\documentclass[final,5p,times,twocolumn]{elsarticle}

 \usepackage{epsfig}

\usepackage{amssymb}

\usepackage{xspace}	
\usepackage{color}
\usepackage{graphicx}
\usepackage{color}

\begin{document}
\begin{frontmatter}

\title{Time Evolution of Temperature Fluctuation in a Non-Equilibrated System}

\author[PHY]{
Trambak~Bhattacharyya
}
\ead{trambak.bhattacharyya@gmail.com}
\author[PHY]{
Prakhar~Garg
}
\ead{prakhar@rcf.rhic.bnl.gov}
\author[PHY]{
Raghunath~Sahoo
}
\ead{Raghunath.Sahoo@cern.ch}

\author[ASTRO]{
Prasant Samantray
}
\ead{prasant.samantray@iiti.ac.in }

\address[PHY]{Discipline of Physics, School of Basic Science, Indian Institute
of Technology Indore, Khandwa Road, Simrol, MP-452020, India}
\address[ASTRO]{Centre of Astronomy, School of Basic Science,
Indian Institute of Technology Indore, Khandwa Road, Simrol, MP-452020, India}

\begin{abstract}
The evolution equation for inhomogeneous and anisotropic temperature 
fluctuation inside a medium is derived within the ambit of Boltzmann
Transport Equation (BTE) for a hot gas of massless particles. Also,
specializing to a situation created after heavy-ion collision (HIC), we 
analyze the Fourier space variation of temperature fluctuation of
the medium using its temperature profile. The effect of viscosity on the
variation of fluctuations in the latter case is investigated and possible
implications for early universe cosmology, and its connection with HICs are also
explored.
\end{abstract}

\end{frontmatter}

\section{Introduction}

Fluctuation is a normal occurrence in physical systems.
Caused by their stochastic nature, the value of certain
observables deviate from their average value, which may be defined over a large
time or over a large number of identically prepared systems (ensembles).
The well-known phenomenon of critical opalescence, for example, is caused due to
fluctuations at all length scales during a second order phase transition.

In high-energy collision experiments, the search for fluctuations of
quantities (like net-charge \cite{jeon2}, \cite{asakawaprl}) over large number
of events are important for searching the critical point \cite{stephanov1} or
tri-critical point \cite{stephanov2} in the quantum chromodynamic (QCD) phase
diagram. The study of particle multiplicity ratio fluctuation \cite{NA49}
is another such example in this context.

Much in the same way as number of particles in a certain region of a system
fluctuates, the everyday examples teach us that the temperature for physical
systems can also fluctuate. Apart from the examples from high-energy collisions
where particle yield has shown the signature of temperature fluctuation
\cite{STAR,PHENIX1,PHENIX2,ALICE,CMS1,CMS2,ATLAS,ALICE2,bediaga,azmi,worku},
there are numerous other situations (like cosmological perturbations in our
expanding universe as sources of temperature fluctuation) where the concerned
system is not in global thermal equilibrium. The temperature, on the contrary
varies with time and space. The temperature fluctuation associated with such
kinds of systems encode transport properties like conductivity, shear viscosity,
rates of chemical reactions etc. As the system evolves, dynamics dictates the
temperature fluctuation until a state of minimum energy, or equilibrium is
attained.

The evolution of the fluctuations has been investigated for systems concerned
with high-energy collisions employing Boltzmann Transport Equation (BTE)
\cite{alametal,stephanov3}. It is now important to study
temperature fluctuation as well as its evolution in such systems as
they can characterize the medium created after high-energy collisions
\cite{sumit1,sumit2} or may give a hint to the QCD critical point
\cite{pgjpg}.

\begin{figure}[h]
\centering
\includegraphics[scale=0.2]{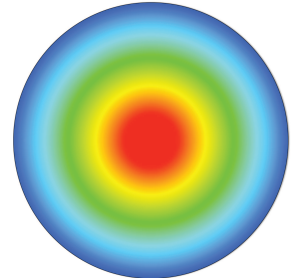}
\caption{(color online) Pictorial representation of hotspots in a medium.}
\label{hotspots}
\end{figure}

A medium with spatially fluctuating temperature can be schematically represented
by Fig. \ref{hotspots}, where, within a large system, we encounter subsystems
with different temperature values. In our present work, we model the time
evolution of temperature fluctuation among these subsystems. At any certain time
slice, we assume that the system comprises of temperature hotspots or zones
evolving with time. This is essentially the assumption of local thermodynamic
equilibrium of matter, where the temperature hotspots are weakly interacting.
The assumption of weakly interacting hotspots is justified by the following - in
any thermal system, the correlation distance may be taken to be the Debye length
($r_D$) which is $\sim (gT)^{-1}$ \cite{bellac}, where $g$ is the coupling and
$T$ is the average temperature. For $g = 0.5$ and $T = 200 $ MeV, $r_D$ is $
\sim 2~fm$. Therefore, for the specific example of the medium created after
Heavy-Ion Collision (HIC), the system radius $r_S>>r_D$ when $T=200$ MeV; and it
can be safely argued that the temperature hotspots are effectively
non-interacting. This in turn implies that particles in a certain temperature
zone hardly affect those in other temperature zones. In fact, (assuming zero
chemical potential) such systems can be represented by a collection of canonical
ensembles \cite{polytherm} with different temperature values. The probability
that a certain member of the ensemble will be having a certain energy at some
instant will depend on the fluctuating temperature values of the collection of
subsystems.

In the present work, we try to find out an evolution equation of
the fluctuation in Boltzmann parameter $\beta~(=1/T)$ with the aid of
Boltzmann Transport Equation in Relaxation Time Approximation (RTA) assuming a
constraint that the observation time is much less than the relaxation time of
the thermal bath. Later on we analyze the same problem for arbitrary observation
time.

Hence, the manuscript is organized as follows. In section \ref{setup} we
begin by considering the BTE and the evolution of $\beta$-fluctuation,
followed by analysis specific to heavy-ion collisions with arbitrary observation
time. We then discuss our results in section 3 where the relative variance of
the Boltzmann $\beta$-parameter will be compared with the similar quantities
extracted from experimental data. Lastly, we conclude by conjecturing possible
connections with early universe cosmology.

\section{The Methodology}
\label{setup}
\noindent In order to gain qualitative insight into the evolution of
temperature fluctuation, we consider an ansatz \cite{dodelson} 
of the particle distribution function $f$ as

\begin{equation}
 f=e^{-\beta p (1 + \Delta \beta)}
\end{equation}

\noindent where we consider a medium with Boltzmann distribution of massless particles 
($p = |\vec{p}| = E$) with average inverse temperature $\beta(t)$, at some time
slice. In the high temperature regime ($\beta E<<1$), that we are
interested in, quantum statistics tend towards Boltzmann distribution. The
average inverse temperature of the system is calculated considering the
arithmetic mean of the distribution of temperature hotspots, {\it i.e.} if there
are $n_i$ hotspots individually characterized by inverse temperatures $\beta_i$,
then the average is calculated as $\frac{\Sigma n_i \beta_i}{\Sigma n_i}$.
Generalizing this to the continuum limit, we add an anisotropic and
inhomogeneous fluctuation function $\Delta \beta(\vec{r},\hat{p};t)$, where
$\hat{p}$ is an unit vector along the direction of motion of particle.
The $\hat{p}$ dependence encodes the anisotropy of the fluctuation. We
now discuss the temporal evolution of the fluctuation with the help of BTE. \\
\\
\noindent The generic form of BTE can be written as:
\begin{equation}
\frac{df}{dt}=\frac{\partial f}{\partial t}+\vec{v}.\vec{\nabla} f+\vec{F}.\vec{\nabla}_p f=\mathcal{C}[f]
\label{bte}
\end{equation}

\noindent where $\vec{v}$ is particle velocity, $\vec{F}$ is any external
force (like gravity) and $\mathcal{C}[f]$ is the collision term encoding the
information about interaction. $\vec{\nabla}_p$ is the momentum-space gradient
operator. For our present case, we assume that the system experiences no
external force, and hence $\vec{F}=\vec{0}$. However, the inhomogeneity in
$\Delta \beta$ still exists. Assuming the $|\Delta \beta| <<1$, we get

\begin{eqnarray}
 f &\approx& \left. e^{-p\beta-p\beta\Delta \beta}\right\vert_{\Delta \beta=0}+
 \left. \beta \frac{\partial}{\partial \beta} \left[{e^{-p\beta-p\beta\Delta \beta}}\right]
 \right \vert_{\Delta \beta=0} 
 \Delta \beta \nonumber\\
 &=& e^{-p\beta}-p\beta e^{-p\beta} \Delta \beta \nonumber\\
 &=& f^{(0)}-f^{(0)} p\beta \Delta \beta
 \label{f}
\end{eqnarray}

\noindent Using $f^{(0)}=e^{-p\beta}$ and putting Eq. (\ref{f}) in Eq. (\ref{bte}), we get

\begin{eqnarray}
\frac{\partial}{\partial t} \left[ e^{-p\beta}-p\beta e^{-p\beta} \Delta \beta \right] 
+ \frac{p^i}{E} \frac{\partial}{\partial x^i} \left[ e^{-p\beta}-p\beta e^{-p\beta} \Delta \beta \right]  \nonumber \\
= - \frac{f-f^{(0)}}{t_{\mathrm{R}}}  
\label{fluceq} 
\end{eqnarray}

\noindent where $v^i=p^i/E$, and we assume the relaxation time approximation for the 
collision term $\mathcal{C}[f]$, with $t_{\mathrm{R}}$ as the relaxation time. Since 
equilibrium distributions are stationary and (in absence of external force) homogeneous, 
the BTE for equilibrium distributions is identically satisfied. In the present scenario, 
we assume the equilibrium distribution function $f^{(0)}$ to be stationary for a time 
duration much longer than the observation time allowed by BTE (but this time
should be much less than $t_R$, within which the distribution changes
appreciably), then

\begin{eqnarray}
\frac{\partial}{\partial t} f^{(0)}+\frac{p^i}{E} \frac{\partial}{\partial x^i} f^{(0)}=0
\end{eqnarray}
\noindent and hence, Eq. (\ref{fluceq}) becomes
\begin{eqnarray}
-p\frac{\partial \Delta \beta}{\partial t} \beta f^{(0)} - \frac{p^i}{E} \beta p f^{(0)} 
\frac{\partial \Delta \beta}{\partial x^i}= \frac{p\beta \Delta \beta f^{(0)}}{t_{\mathrm{R}}}
\label{fluceq1}
\end{eqnarray}
if we assume the average inverse temperature to be changing very slowly with
time.
\\
 
\noindent Expressing $\Delta \beta(\vec{r},\hat{p};t)$ in terms of its Fourier Transform

\begin{eqnarray}
 \Delta \beta(\vec{r},\hat{p};t) = \int d^3k \Delta \beta_k(t)
e^{i\vec{k}.\vec{x}}
\end{eqnarray}

\noindent where we denote $\Delta \beta(\vec{k},\hat{p};t)\equiv\Delta
\beta_k(t)$ for simplicity.
Eq. (\ref{fluceq1}) becomes

\begin{eqnarray}
-p\beta \frac{\partial \Delta \beta_k(t) }{\partial t} -
\frac{p\beta}{t_{\mathrm{R}}} \Delta \beta_k(t) -
i\beta \frac{|p|}{E} p k \mu \Delta \beta_k(t) = 0 \nonumber\\
\frac{\partial \Delta \beta_k(t) }{\partial t} = -\left[i\frac{|p|}{E} k \mu +
\frac{1}{t_{\mathrm{R}}}\right]
\Delta \beta_k(t)
\label{flucevoeq}
\end{eqnarray}

\noindent where $\hat{k}.\hat{p}=\mu=\mathrm{cos}\theta$ ($\theta$ is the
angle $\hat{k}$ makes with $\hat{p}$). The solution of Eq. (\ref{flucevoeq}) is
then given by (see \cite{sarwaralam}, for example, for similar equation in
context of energy density fluctuation.)
 
\begin{equation}
 \Delta \beta(\vec{k},\hat{p};t) = \Delta \beta(\vec{k},\hat{p};t^0) e^{-i \frac{|p|}{E} k \mu (t-t^0)} 
 e^{-\frac{t-t^0}{t_{\mathrm{R}}}}
 \label{tempflucevoeq}
\end{equation}

\noindent We can simplify Eq. (\ref{tempflucevoeq}) by assuming an isotropic
fluctuation profile. Thus, assuming $|\vec{p}|=E$ we get a simplified expression
for the temporal evolution of fluctuation for a medium of massless particles. We
further average over the whole solid angle $\Omega$ subtended by $\hat{p}$. Here
$\vec{k}$ is a constant vector assumed to be directed along the $z$-axis. Hence,
the averaged fluctuation becomes, 

\begin{eqnarray}
\Delta \beta_{\mathrm{av}} (\vec{k};t) &=& \Delta \beta (\vec{k};t^0) e^{-\frac{t-t^0}{t_{\mathrm{R}}}}
\frac{1}{4\pi} \int_{\Omega} e^{-ik\mu(t-t^0)} d\Omega \nonumber\\
&=& \Delta \beta (\vec{k};t^0) e^{-\frac{t-t^0}{t_{\mathrm{R}}}}
\frac{1}{4\pi} \int_{-1}^{1} d\mu e^{-ik\mu(t-t^0)} \int_{0}^{2\pi} d\phi \nonumber\\
\Delta \beta_{\mathrm{rel}} (\vec{k};t) &=& 
\frac{\Delta \beta_{av} (\vec{k};t)}{\Delta \beta (\vec{k};t^0)} \nonumber\\
&=&  e^{-\frac{t-t^0}{t_{\mathrm{R}}}}\frac{\mathrm{sin} k(t-t^0)}{k(t-t^0)}
\label{relfluc}
\end{eqnarray}

\begin{figure}[h]
\centering
\includegraphics[scale=0.6]{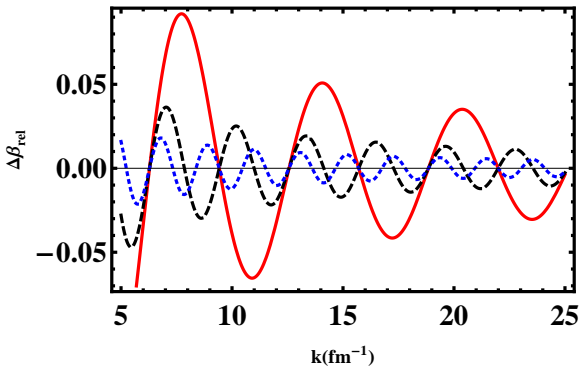}
\caption{(color online) Variation of $\Delta \beta_{\mathrm{rel}} (\vec{k};t)$
with $k$. Red (solid): $(t-t^0) =1 ~fm$, Black (dashed):
$(t-t^0) = 2~fm$ , Blue (dotted): $(t-t^0) = 3~ fm$ for $t_{\mathrm{R}} = 3
~fm$.}
\label{TFlucBTETime}
\end{figure}

\begin{figure}[h]
\centering
\includegraphics[scale=0.6]{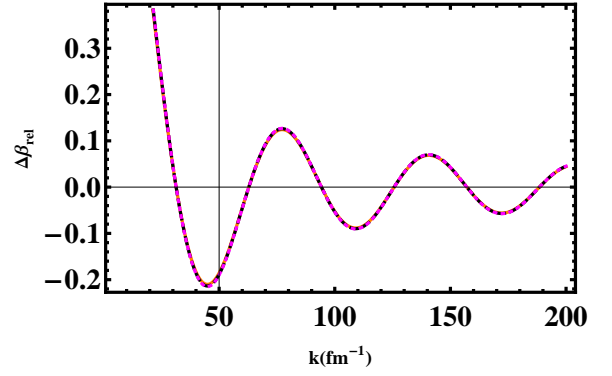}
\caption{(color online) Variation of $\Delta \beta_{\mathrm{rel}} (\vec{k};t)$
with $k$. Orange (solid): $t_{\mathrm{R}} = 3 ~fm$, Black
(dashed): $t_{\mathrm{R}} = 6 ~ fm$, Magenta (dotted) $t_{\mathrm{R}} = 9~ fm$
for $(t-t^0) = 0.1~fm $.}
\label{TFlucBTERelTime}
\end{figure}

From Eq. (\ref{relfluc}), we can infer that the relative fluctuation
$\Delta \beta_{\mathrm{rel}} (\vec{k};t)$ is monotonically decreasing. In Fig.
\ref{TFlucBTETime} as well as in Fig. \ref{TFlucBTERelTime}, we provide the
plots depicting the parametric Fourier space variation of the $\Delta
\beta_{\mathrm{rel}} (\vec{k};t)$ with time ($t-t^0$) and
relaxation time ($t_{\mathrm{R}}$) respectively. The reliability of the
variations shown in figures is governed by the constraint that the observation
time must be much less than the time taken by the distribution function to
change appreciably \cite{balescu}, {\it i.e.} the relaxation time $t_R$.

\begin{equation}
(t-t^0)<<t_{\mathrm{R}}
\label{timeupperlimit}
\end{equation}
According to our earlier assumption about very slow variation of $\beta$ with
time, ($t-t^0$) should also be such a time-interval within which we can assume
almost constant temperature.
\\

\noindent We observe in Fig. \ref{TFlucBTETime} that the relative fluctuations
die down with time. Additionally, the soft modes of fluctuations, or in other
words, fluctuations at larger distances towards the periphery of the medium, are
large. In Fig. \ref{TFlucBTERelTime}, we observe no
modification of fluctuation with increasing $t_{\mathrm{R}}$ when
$(t-t^0)<<t_{\mathrm{R}}$.

We have thus solved the evolution equation for the $\beta$-fluctuation in the
Fourier space using Boltzmann Transport Equation. However, the generality of our
calculation is limited by the upper bound in Eq. (\ref{timeupperlimit}).
Consequently, Eq. (\ref{tempflucevoeq}), which assumes very slow variation of
temperature, cannot be applied to certain cases involving arbitrarily large
observation times within which temperature changes appreciably. With the end to
study a more realistic situation, we can consider the temperature profiles of a
evolving medium at different time slices which are arbitrarily separated. After
quantifying the inverse temperature fluctuation, we can find out the inverse
temperature fluctuation at every time-instant and will try to observe their
variation at different stages.

As an example, we have chosen the radially varying temperature profile of a
viscous medium created by heavy-ion collisions from Ref. \cite{baiertempprof}.
We can characterize the temperature profile of a viscous medium shown in Ref.
\cite{baiertempprof} by the following function.

\begin{eqnarray}
T_M(r,t)=\frac{T_0(t)}{e^{a(t)\left(\frac{r}{r_0}-1\right)}+1}
\label{TempProfBaier}
\end{eqnarray}

\noindent where at $r=r_0$, $T_M(r)=T_0/2$; and $T_M(r)\approx T_0$ at $r=0$ and $a(t)$ is 
a parameter which fixes how sharply the function drops down.
From Eq. \ref{TempProfBaier},
writing $\beta_M=1/T_M$ we get

\begin{equation}
\beta_M(r;t)=\beta_0(t)\left(e^{a(t)\left(\frac{r}{r_0}-1\right)}+1\right)
\label{betadist}
\end{equation}

\noindent where $r$ denotes the radial distances of the zones from the centre
of the medium. Using Eq. (\ref{betadist}), we can generate $\{\beta_{M}\}$ -- a
collection of $\beta_{M}$ values. Given the collection, we can now define an
average $\beta_M$ value $\langle \beta_M \rangle =\beta(t)$ at a certain instant
$t$ and can define a fluctuation $\Delta\beta(r,t)$ as

\begin{eqnarray}
\Delta\beta(r,t)&=&\beta_M(r,t)-\beta(t) \nonumber\\
&=&\beta_0(t)e^{ a(t) \left( \frac{r}{r_0} - 1 \right)}+\delta \beta(t)
\end{eqnarray}

\noindent where $\delta \beta(t)=\beta_0(t)-\beta(t)$. The Fourier Transform $\Delta\beta(k;t)$ 
now becomes

\begin{eqnarray}
\Delta\beta(k;t) &=&
 \Delta\beta_k \nonumber\\
 &=&\frac{2\beta_0(t)}{(2\pi)^2k} \int_{0}^{R}
 e^{ a(t) \left( \frac{r}{r_0} - 1 \right)} r \mathrm{sin}(k~r) dr  + \delta
\beta(t) \delta(\vec{k}) \nonumber\\
 \label{TempFlucFT}
\end{eqnarray}

\noindent where $R$ is the system size and $\delta(\vec{k})$ is the Dirac delta
function.

\begin{figure}[!htb]
\minipage{0.7\textwidth}
\includegraphics[scale=0.58]{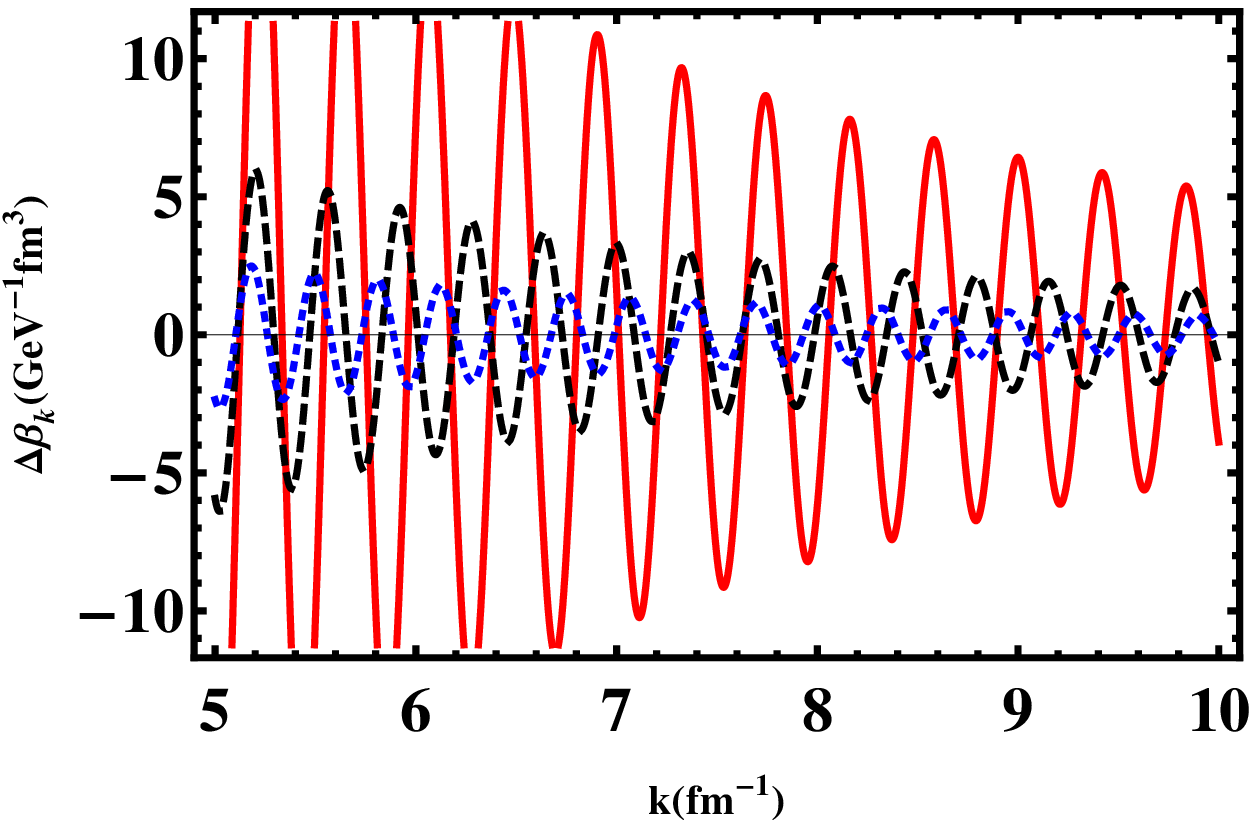}
\endminipage\hfill
\minipage{0.7\textwidth}
\includegraphics[scale=0.58]{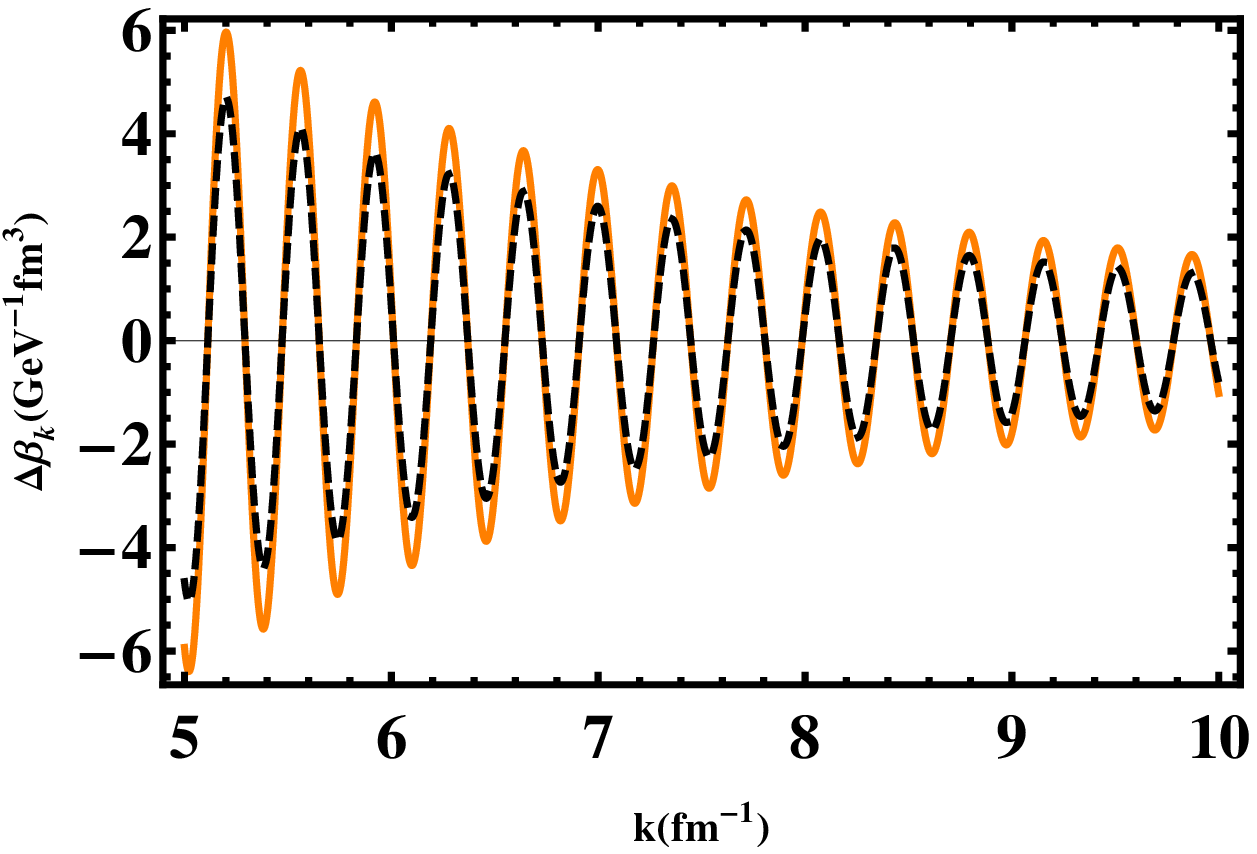}
\endminipage\hfill
\caption {(color online) Variation of inverse temperature fluctuation (Eq.
\ref{TempFlucFT}) in a viscous medium with $k$. (upper panel) Red(solid):
$\tau=2.2$ fm/c, Black(dashed): $\tau=5.1$ fm/c, Blue(dotted): $\tau=9.1$ fm/c.
$\eta/s=0.08$ for all the figures. (lower panel) Orange(solid): $\eta/s
= 0.08$, Black(dashed): $\eta/s = 0.3$. at $\tau=5.1$ fm/c }
\label{TFluc}
\end{figure}

\section{Results and Discussion}

As seen from Eq. (\ref{TempFlucFT}), the soft modes of $\beta$-fluctuation
become dominant implying that towards the periphery (at large system radius),
the fluctuation is greater. The variation of the inverse temperature fluctuation
in the momentum space is shown in the upper panel of Fig. \ref{TFluc}. The lower
panel of Fig. \ref{TFluc} shows the variation of fluctuation for different
viscosities of the medium. As intuitively expected, higher viscosity favours
lower fluctuations. 

In the previous section, we have already defined the average $\beta(t)$ and
fluctuation $\Delta \beta$ with the help of the set $\{\beta_{M}\}$. With the
aid of the same set we can now define a relative $\beta$-fluctuation.

\begin{equation}
\frac{\langle \beta_M^2 \rangle-\langle \beta_M \rangle^2}{\langle \beta_M
\rangle^2} = \mathcal{R}_{\beta}
\label{rbeta}
\end{equation}

\noindent Using the $\beta_0$, $a$ and $r_0$ values as tabulated in Table
\ref{parvalue}, we compare the $\mathcal{R}_{\beta}$ in the
system produced in HICs at different stages of its evolution with the help of
Eq. (\ref{betadist}).

We observe that within any arbitrary choice of radius shell the relative
fluctuations die down with time. For demonstration, we have chosen the 
shell ranging between the radii 14 {\it fm} to 15 {\it fm} in Table
\ref{parvalue}. But, our observation remains unaltered for any other
shell.

In Table \ref{relflucvisco}, we show the change in $\mathcal{R}_{\beta}$
with viscosity (for a radius shell ranging between 14 $fm$ to 15 $fm$). With
increasing viscosity, the relative fluctuation decreases, thereby leading to
lower $\mathcal{R}_{\beta}$ values.

\begin{table}[h]
\caption{Values of parameters extracted from the temperature profile shown in
 Ref. \cite{baiertempprof} using Eq. \ref{TempProfBaier} and the
 $\mathcal{R}_{\beta}$ values using Eq. \ref{rbeta} at different $\tau$ with
 $\eta/s$=0.08.}
\begin{center}
 \begin{tabular}{ |c|c|c|c|c| }
 \hline
$\tau$(fm/c) & $\beta_0$(GeV$^{-1}$) & $a$ & $r_0$(fm) &
$\mathcal{R}_{\beta}$ \\
\hline
2.2 & 3.45 & 5.99  & 7.96 & 0.047\\
\hline
5.1 & 4.55 & 3.42  & 8.41 & 0.011\\
\hline
9.1 & 5.56 & 1.91  & 8.71 & 0.002\\
\hline
\end{tabular}
\end{center}
\label{parvalue}
\end{table}

 \vspace{0.03cm}

\begin{table}[h]
\caption{Relative fluctuations in $\beta$ at $\tau$=5.1 fm/c with change of
 viscosity.}
\begin{center}
\begin{tabular}{ |c|c| }
\hline
$\eta/s$ & $\mathcal{R}_{\beta}$ \\
\hline
0.08 & 0.012\\
\hline
0.3 & 0.011\\
\hline
\end{tabular}
\end{center}
\label{relflucvisco}
\end{table}

\begin{table}[h]
\begin{center}
\caption{Comparison of $\mathcal{R}_{\beta}$ at the boundary as obtained from
Eq. \ref{TempProfBaier} with the $(q-1)$ value obtained from experiment
\cite{TBW}.}
\vspace{0.3cm}
\begin{tabular}{ |c|c| }
\hline
$\mathcal{R}_{\beta}$ & $(q-1)$  \\
\hline
0.01 & 0.018$\pm$ 0.005 \\
\hline
\end{tabular}
\end{center}
\label{rbetaqminus1}
\end{table}

%

As it turns out, the multiparticle production processes in high-energy
electron-positron \cite{bediaga}, hadronic and heavy-ion collisions
\cite{TsallisInHIC1,TsallisInHIC2,TsallisInHIC3,TsallisInHIC4,TsallisInHIC5,
TsallisInHIC6,TsallisInHIC7,TsallisInHIC8} are quite accurately characterized by
a Tsallis entropic parameter $q$ \cite{Tsallis}, which is similar to
$\mathcal{R}_{\beta}$, and lies typically in the range $1 < q < 1.2$
\cite{beck1} in context of high-energy collisions. Here, we would like to
briefly mention some recent works done by the authors in \cite{Tsallis,Wilk}
connecting the $q$-parameter and the temperature fluctuation or {\it
non-extensivity} of thermal systems. The non-extensive nature is manifested
once we find out that the simple addition of entropies ($S$) of two sub-parts
($A$ \& $B$) of a bigger system $C$ does not give the entropy of the system
$C$. Rather, $S(C)=S(A)+S(B)+(1-q)S(A)S(B)$, where $q$ measures the degree
of deviation from the additive domain. This leads to a proposal of modification
of the usual Boltzmann-Gibbs formula to

\begin{equation}
G_{q}(x) = \left[1 + (q-1)\beta E \right]^{\frac{-1}{q-1}}
\end{equation}

\noindent As $q  \rightarrow 1$, $G_{q}(x) \rightarrow e^{-\beta E}$, and we
recover the usual Boltzmann-Gibbs formula. Therefore, $q$ has also been dubbed
as the non-extensivity parameter in the literature. In an elegant exposition of
the same, Wilk \cite{Wilk} deduced that

\begin{eqnarray}
G_{q}(x) &=& \left[1 + (q-1)\beta E \right]^{\frac{-1}{q-1}} \nonumber \\
&=& \int^{\infty}_{0}~e^{-\beta' E} f(\beta')d\beta'
\end{eqnarray}

\noindent where the distribution function $f(\beta')$ is the usual chi-squared
function given by

\begin{equation}
f(\beta') = \frac{\alpha \beta}{\Gamma(\alpha)} \left(\frac{\alpha \beta}
{\beta'} \right)^{\alpha - 1} \exp {\left(-\frac{\alpha \beta}{\beta'}
\right)}
\end{equation}

\noindent where $\alpha = \frac{1}{q-1}$. With respect to the above chi-squared
distribution, we have the mean value $\langle \beta' \rangle = \beta$, and
also the relative variance as

\begin{equation}
\frac{\langle \beta'^2 \rangle- \langle \beta' \rangle^2}{ \langle \beta'
\rangle ^2} = q-1
\label{qminus1}
\end{equation}
We can therefore make a correspondence between the
$\mathcal{R}_{\beta}$ defined in Eq. \ref{rbeta} and the $(q-1)$ defined in 
Eq. \ref{qminus1}. The non-zero values of $q$ are associated not only
with the relative $\beta$ fluctuation in the system, but also with that during
the hadronization process \cite{Wilk,beck,beckcohen}. This also gains
significance in the context of the QCD phase diagram and the search for critical
point which, in fact, may be associated with large fluctuations in the $q$ value
\cite{pgjpg}. Non-extensivity of any thermodynamic system is
invariably linked to temperature fluctuation, and hence, the heat
capacity/specific heat of the system. However, whether the converse statement
holds is yet to be answered.
\\
\noindent The $q$ values for systems produced in high-energy
collisions can be obtained \cite{TBW} by fitting the experimentally
observed particle spectra. Assuming an average freeze-out time of $\sim 9$ fm
we can study the temperature profile at $\tau=9.1$ fm to
compare the $\mathcal{R}_{\beta}$ values with the experimentally observed
$(q-1)$ values under similar conditions \cite{TBW}. According to the Table
\ref{rbetaqminus1},  
the $\mathcal{R}_{\beta}$ value ($\sim0.01$) at the system boundary is
comparable with the experimentally obtained value ($\sim0.018\pm0.005$ for
0-10$\%$ central HICs at RHIC with $\sqrt{s}_{NN}$=200 GeV\cite{TBW}). This 
observation re-emphasizes the relationship between temperature fluctuation and
the $q$ parameter \cite{Wilk}.
\\
Additionally, some comments about connecting certain observables in HICs with
the theory of cosmological perturbations are in order. Cosmological anisotropies
reflect the energy content of the universe. The universe starts off as radiation
dominated, changes over to being matter dominated, and is eventually
conjectured to be purely governed by the cosmological constant ($\Lambda$)
\cite{dodelson}. WMAP \cite{wmap} and Planck \cite{planck} both provide a fairly
precise representation of the energy distribution at our current epoch - via
physical quantities like $\Omega_{matter} $, $\Omega_{\Lambda}$,
$\Omega_{baryons} $, the acoustic scale, Hubble constant, neutrino fraction,
reionization optical depth and other derived quantities. Since the anisotropies
in temperature fluctuation are all time dependent, in later epochs these
fluctuations would die down, and theoretically one should expect a flat power
spectrum in the infinite future. However, authors in
\cite{Bernui-Tsallis-Villela} have conjectured that the temperature fluctuation
of our universe can be satisfactorily explained by the {\it modified
Boltzmann-Gibbs} formula with $q = 1.045 \pm 0.005$. This is quite remarkable
since 
the similar $q$-value that fits heavy-ion collision data also fits the data for
cosmological fluctuations. This points to deep similarities between the physics
of cosmic microwave background (CMB) radiation anisotropies and the flow
anisotropies in relativistic heavy-ion collision experiments (RHICE). A relevant
theoretical question would be - is the surface of last scattering for CMB
radiation similar to the freeze-out surface in RHICE? This is a question we
reserve for future work. \\

\noindent {\bf Acknowledgements}: The authors would like to thank Prof.
Jan-e Alam, Dr. Amaresh Jaiswal, Dr. Moumita Aich and Golam Sarwar for
fruitful discussions. TB and PG acknowledges the financial support by DST,
Govt. of India.

\end{document}